# Reactive magnetron sputtering of tungsten target in krypton/trimethylboron atmosphere

Martin Magnuson, Lina Tengdelius, Fredrik Eriksson, Mattias Samuelsson, Esteban Broitman,† Grzegorz Greczynski, Lars Hultman, and Hans Högberg

*Department of Physics, Chemistry, and Biology (IFM) Linköping University, SE-581 83 Linköping, Sweden*
†*Present address: SKF Research and Development Center, 3439 MT Nieuwegein, Netherlands*

2019-06-19

*Corresponding author: martin.magnuson@liu.se

## Abstract

W-B-C films were deposited on Si(100) substrates held at elevated temperature by reactive sputtering from a W target in Kr/trimethylboron (TMB) plasmas. Quantitative analysis by X-ray photoelectron spectroscopy (XPS) shows that the films are W-rich between ~ 73 and ~ 93 at.% W. The highest metal content is detected in the film deposited with 1 sccm TMB. The C and B concentrations increase with increasing TMB flow to a maximum of ~18 and ~7 at.%, respectively, while the O content remains nearly constant at 2-3 at.%. Chemical bonding structure analysis performed after samples sputter-cleaning reveals C-W and B-W bonding and no detectable W-O bonds. During film growth with 5 sccm TMB and 500 °C or with 10 sccm TMB and 300-600 °C thin film X-ray diffraction shows the formation of cubic 100-oriented $WC_{1-x}$ with a possible solid solution of B. Lower flows and lower growth temperatures favor growth of W and $W_2C$, respectively. Depositions at 700 and 800 °C result in the formation of $WSi_2$ due to a reaction with the substrate. At 900 °C, XPS analysis shows ~96 at.% Si in the film due to Si interdiffusion. Scanning electron microscopy images reveal a fine-grained microstructure for the deposited $WC_{1-x}$ films. Nanoindentation gives hardness values in the range from ~23 to ~31 GPa and reduced elastic moduli between ~220 and 280 GPa in the films deposited at temperatures lower than 600 °C. At higher growth temperatures the hardness decreases by a factor of 3 to 4 following the formation of $WSi_2$ at 700-800 °C and Si-rich surface at 900 °C.

**Keywords:** W-B-C films, reactive magnetron sputtering, trimethylboron, nanoindentation, X-ray photoelectron spectroscopy, thin film X-ray diffraction, Scanning Electron Microscope

# 1. Introduction

Tungsten carbide and the δ-WC phase [1] with a hexagonal stacking sequence (Strukturbericht designation $B_h$) is a technologically important material best illustrated from its use on cemented carbide tools for metal machining. Favorable to applications, the δ-WC property envelope includes: a melting point of 2870 °C, Vickers hardness (HV) of 22 GPa, Young's modulus of elasticity 620-720 GPa, and good oxidation resistance [2]. If δ-WC could be grown as a thin film material the number of applications would increase substantially by exploiting for instance the carbide's low resistivity of 17-22 μΩ·cm [2]. A low resistivity in combination with the previously defined properties is thus favorable in high-temperature Schottky contacts [3]. However, the literature is clear on the difficulties in depositing δ-WC by chemical vapor deposition (CVD) or physical vapor deposition. Presently, CVD seems to be the most promising technique for growth of films containing the phase [4] [5] [6] [7] [8]. The high temperatures typical for these studies with 1200 °C [4], 950 °C [6], and 900 °C [7] [8] as well as the restriction to W(110) substrates [5] limits the number of applications and calls for alternative deposition techniques. In addition, competing phases are frequently present in the deposited films seen from the tungsten rich carbide $W_2C$ and the cubic and substoichiometric γ-WC with NaCl-type structure (Strukturbericht designation B1) [1], henceforth referred to as $WC_{1-x}$. The properties for the metastable $WC_{1-x}$ are, however, not as favorable as those for δ-WC seen from carbon vacancies with x~0.6 [1]. In carbides with B1 structure such as $WC_{1-x}$ carbon vacancies will result in lower hardness [2] [9].

Magnetron sputtering offers the possibility for low-temperature growth and the literature shows that WC films have been deposited from WC targets, [10] [11] [12] [13], W and C sources [14] [14] [15] and of particular interest for this study by reactive sputtering. The most applied gaseous reactants have been $C_2H_2$ (acetylene), [11] [16] [17] [18] [19] [20] and $CH_4$ (methane) [21] [22], but with report from reactive processing from other hydrocarbons such as ethene ($C_2H_4$) [12] and benzene ($C_6H_6$) [23]. As experienced from CVD, there are difficulties in depositing the δ-WC phase with only a few reports on reactive sputtering of phase-pure films, using alloying with $N_2$ [24] or growth on carbon substrates at temperatures > 900 °C [12].

Thus, the prospect of depositing single-phase δ-WC films by sputtering seems difficult. For this task, alloying by a third element such as B constitutes a viable route to improve the properties of $WC_{1-x}$ films by filling C vacancies with B atoms. Liu *et al.* have reported reactive sputtering from a $WB_2$ target in $C_2H_2$ containing plasmas for deposition of W-B-C films [25]. The study shows that the composition of the films can be varied at low flows of $C_2H_2$, but with deposition of almost pure carbon films when the $WB_2$ target is poisoned by $C_2H_2$. In two publications, Alishahi *et al.* [26] and Debnárová *et al.* [27] investigated growth of W-B-C films from magnetron sputtering of W, and $B_4C$ targets, in combination with pulsed sputtering of a C target. The applied experimental set-up with three separate sources allows for growth of films in a with a wide range of compositions and properties as reported by the authors but is less practical for industrial applications.

An alternative and more industrially compatible route is to apply boron as a gaseous reactant and then preferably as a single precursor containing both carbon and boron. In a seminal study, Lewis *et al.* [28] compared trialkylboron triethylboron [TEB, $B(C_2H_5)_3$], trimethylboron (TMB, $B(CH_3)_3$) and tributylboron [TBB, $B(C_4H_9)_3$] and suggested that TEB was suitable for depositing boron carbon films by CVD. In addition, the authors have successfully applied TEB [29] [30] [31] for growth of epitaxial $sp^2$-BN films by thermally activated CVD. The fact that TMB is a gas at standard pressure and temperature conditions, while TEB is a liquid and TBB is a solid makes TMB the easiest trialkylboron to integrate in a sputtering process. In addition, TMB exhibits the highest boron-to-carbon of the organoboranes with 1:3 compared to 1:6 in TEB and 1:12 in TBB, which allows for studying boron-rich W-B-C films.

To investigate the possibility of integrating a common precursor from CVD in magnetron sputtering, we reactively sputter a W target with TMB as a single gaseous reactant for growth of W-B-C films. The flow of TMB is varied for deposition of W-rich films and with growth carried out without external heating (room temperature, RT) to 900 °C.

## 2. Experimental details

The W-B-C films were deposited in an ultra-high vacuum system (base pressure $2 \times 10^{-6}$ Pa) on Si(100) substrates by reactive sputtering of a 3 inch circular W target in a Kr plasma (99.998%) using TMB gas with a purity of 99.99% from Voltaix Inc., High Springs, FL, USA as precursor. Krypton was applied as sputtering gas instead of Ar to reduce the probability of back-scattered neutrals that will introduce stresses in the films. All films were deposited for 5 min without substrate rotation, using a sputtering current of 900 mA and with the substrates held at floating potential. In our depositions the TMB gas was distributed close to the substrate table through a gas pipe mounted in the deposition chamber and with the Kr gas was distributed in a separate line at a flow of 82 sccm in all depositions. This set-up was chosen to minimize poisoning of the W target by TMB. The substrate was mounted ~5 cm from the end of the gas pipe and with the substrate positioned directly in line-of-sight above the magnetron at a distance of 7 cm. The substrate was heated by a heating stage positioned directly above the substrates. For additional information on the applied deposition system the reader is referred to ref. [32]. For films deposited with 10 sccm TMB in the plasma the deposition temperature was investigated from RT, 100, 200, 300, 400, 500, 600, 700, 800, and 900 °C. At 500 °C the TMB content in the plasma was varied from 1, 2.5, 5, 7.5, and 10 sccm. The addition of TMB to the plasma caused a slight increase of the total pressure from 0.53 Pa for sputtering without TMB in the plasma to 0.6 Pa at the highest applied flow of TMB as well as increased voltage on target as dependent on the amount of TMB in the plasma. The pronounced poisoning of the target seen from an increasing target voltage from ~450 V to ~600 V limited the applied flow of TMB to 10 sccm in this study. Prior to deposition, the substrates were ultrasonically degreased for 5 min in trichloroethylene, followed by 5 min in acetone, and finally 5 min in isopropanol.

The chemical composition and the chemical bonding structure of the films was assessed by X-ray photoelectron spectroscopy (XPS). The instrument used was an AXIS Ultra DLD from Kratos Analytical, with monochromatic Al Kα radiation (hν = 1486.6 eV), operated at a base pressure of $1.5 \times 10^{-7}$ Pa and with the X-ray anode at 225 W. The binding energy scale was calibrated by setting the position of the Fermi edge of a sputter-cleaned Ag sample to 0.0 eV [33] resulting in the position of the Ag $3d_{5/2}$ core-level peak of 368.30 eV [34]. To remove adsorbed contaminants following air exposure, the samples were sputter-cleaned for 180 s with 4 keV Ar-ions incident at an angle of 70° with respect to the surface normal. Casa XPS software (version 2.3.16) was used for quantification, with elemental sensitivity factors supplied by Kratos Analytical Ltd. The confidence level of XPS is typically around ±5 %. X-ray diffraction (XRD) θ/2θ scans were used to evaluate the structural properties of the films using a Philips PW 1820 Bragg-Brentano diffractometer, equipped with a Cu anode X-ray tube (Cu $K_\alpha$, λ=1.54 Å) operated at 40 kV and 40 mA. Additional XRD scans were performed in grazing incidence (GI) geometry with an incidence angle of α=2°, using a Panalytical Empyrean diffractometer in a parallel beam setup with a line focus Cu-$K_\alpha$ X-ray source operating at 45 kV and 40 mA.

The primary beam was conditioned using an X-ray mirror and a ½° divergence slit, and in the secondary beam path a 0.27° parallel plate collimator was used together with a PIXcel detector in 0D mode for data acquisition. XRD pole figure measurements were performed using the same Empyrean diffractometer although now with a point-focused copper anode source and an X-ray lens with 2x2 mm$^2$ crossed-slits on the incident beam side to reduce the effect of defocussing when tilting the sample. The detector position was fixed at specific diffraction angles, corresponding to diffraction from $WC_{1-x}$ {200}, {220}, and {222} family of planes, respectively. The pole figure measurements were performed in 5°-steps with azimuthal rotation $0° \leq \phi \leq 360°$ and tilting $0° \leq \psi \leq 85°$ to determine the orientation distribution of the crystals.

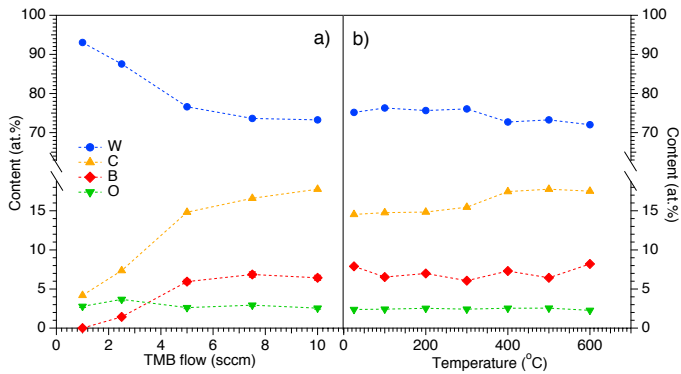

**Figure 1:** Elemental compositions for W-B-C films deposited at 500 °C and with TMB flows of 1, 2.5, 5, 7.5 and 10 sccm in (a) and for films deposited at RT, 100, 200, 300, 400, 500, and 600 °C and with 10 sccm TMB in the plasma in (b).

Film morphologies and thicknesses were investigated using cross-sectional scanning electron microscopy (SEM). The instrument was a LEO 1550 Gemini SEM and the applied acceleration voltage was 10 kV. The nanomechanical properties of the films were measured by quasi-static displacement-controlled nanoindentation tests, using a Hysitron Triboindenter model TI950. Hardness (H) and reduced elastic modulus ($E_r$), were measured with a Berkovich diamond probe and calculated according to the method proposed by Oliver and Pharr [35]. The tip area function was calibrated using a fused silica sample. The thermal drift was compensated prior to each measurement, and no pile-up corrections were necessary. For each test, a total of 12 indents with a spacing of 10 μm were averaged to determine the mean value and standard deviations of H and $E_r$.

The "rule of thumb" for nanoindentations, stating that the indenter should not penetrate more than 10% of the total film thickness, has been shown to fail in many materials (see [36] and references therein). In some of our W-B-C films, the maximum penetration depth $h_{max}$ was calculated using the graphical method explained in the standard ISO 14577 Part 4 [37]. This was validated by performing numerous indentations in one of the harder films at different penetration depths to conclude that the displacement-controlled nanoindentation experiments should be carried out in our samples at $h_{max} = 70$ nm indentation depth.

To be able to relate HV reported by others to our nanoindentation values given in GPa, an approximate conversion of HV values to GPa were made by multiplying the Vickers hardness values by 0.009807, i.e. converting kg/m$^2$ to Pa. The exact conversion from a Vickers hardness into the nanoindentation hardness values needs a geometrical factor correction which is ~0.927 for a perfect Berkovich diamond [36]. It should be noted that the relation is correct only for materials that deform fully plastically during the indentation. If the material has an elastic recovery, the relation is not valid because the Vickers indentation hardness is calculated using the deformed area after indentation, and in the nanoindentation method the hardness is defined by using the deformed area at maximum applied load [36]. To compare reported values on elastic modulus (E) from the literature with our measured $E_r$ values, we applied the formula; $1/E_r=((1-v^2)/E)+((1-v_i^2)/E_i))$ from ref. [36] and assumed: a Poisson's ratio for the sample

ν=0.25, a Poisson's ratio for the indenter of ν=0.07, and the elastic modulus of the indenter to be $E_i$=1140 GPa.

## 3. Results and Discussion
*3.1 Composition and chemical bonding structure*

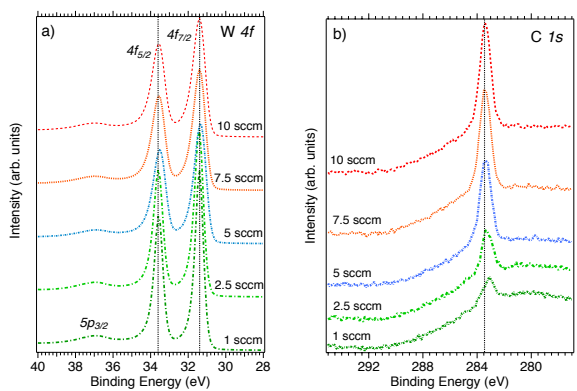

**Figure 2:** XPS spectra from W *4f* in a) and C *1s* in b) of the W-B-C films deposited at 500 °C and with TMB flows from bottom to the top: 1, 2.5, 5, 7.5 and 10 sccm. Literature binding energies for W-W [38] and C-W [24,39] are indicated by the vertical dotted lines in a) and b), respectively.

Figure 1a shows the elemental composition obtained from the quantitative analysis of XPS spectra for W-B-C films deposited with TMB flows of 1, 2.5, 5, 7.5, and 10 sccm at 500 °C. As observed, all films are metal-rich and where the W content decreases from ~ 93 at% to ~ 73 at% from with increasing TMB content in the plasma. As expected, the C and B contents in the films increase with increasing TMB flows from ~ 4 to ~ 18 at.% for C and ~ 0 to ~ 7 at.% for B. The C-to-B ratio is 3:1 for the film deposited with 10 sccm TMB i.e., the same as in the applied TMB precursor, while films deposited with 5 and 7.5 sccm TMB show a more boron rich composition with C/B~2.5, but close to a 3:1 ratio. The film deposited with 1 sccm TMB is carbon rich as evident from C/B~5, which is probably due to uncertainties in determining the B content given the low intensity of the B *1s* peak in XPS. The O content is nearly constant at 2-3 at.%.

Note that all layers were exposed to air before XPS measurements which resulted in a few nm thick native oxides at the surface. Prior to XPS analyses all samples were cleaned with Ar ion beam. Hence, the oxygen detected in the films can have two sources: (1) redeposited atoms following the sputter-etch, and (2) implanted atoms resulting from a forward momentum transfer during Ar sputter-cleaning. Figure 1b shows that the W, C, B, and O content remain nearly constant for films deposited at RT, 100, 200, 300, 400, 500, and 600 °C and with 10 sccm TMB in the plasma. The W content is between 75 to 76 at.% for films deposited at RT to 300 °C and between 72 to 73 at.% for films deposited at 400, 500 and 600 °C. At higher deposition temperatures, Si is detected in the films (not shown) suggesting a reaction with the Si(100) substrate, which is further discussed with the XRD results. Carbon shows an opposite trend compared to W with about 15 at.% below 300 °C and 17-18 at.% above 300 °C, while no clear trend is found for B that varies between 6-8 at.%. We suggest that the difference in W and C content in the films is due to a more efficient dissociation of the TMB precursor at higher temperatures, resulting in a slightly carbon-rich composition in the temperature range from 300 to 600 °C. Similar as for the films deposited with different flows of TMB in the plasma the oxygen content is constant at 2-3 at.%.

Figure 2a and 2b show W *4f* and C *1s* core-level XPS and Fig. 3a and 3b show B *1s* and O *1s* for W-B-C films deposited with TMB flows of 1, 2.5, 5, 7.5, and 10 sccm at 500 °C. The W $4f_{7/2}$ and $4f_{5/2}$ peaks in Fig. 2a, indicated by the vertical dotted lines, are located at 31.4 eV and 33.6 eV, respectively, with a spin-orbit splitting of 2.2 eV. These values are identical to those reported in the literature for metallic W [38], thus supporting the metal-rich composition determined for the reactively sputtered films. Measurements by Greczynski *et al.* on sputtered $W_2C$ films with a hexagonal crystal structure give a binding energy of 31.8 eV for the $W4f_{7/2}$ peak [39], *i.e.* a chemical shift of 0.4 eV to higher binding energy compared to in our films, probably due to the higher C content in the analyzed $W_2C$ films. There is no double-peak around 32.8-35.8 eV for W-O bonding [34]. Instead, the low intensity structure located at 36.8 eV originates from W *5p$_{3/2}$* states.

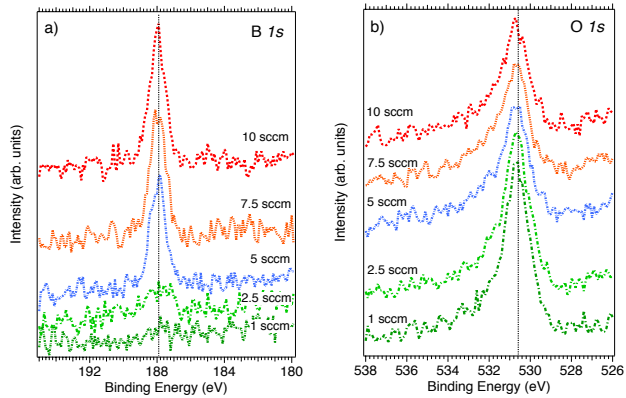

**Figure 3:** B *1s* in a) and O *1s* in b) of XPS spectra from the W-B-C films deposited at 500 °C and with TMB flows from bottom to the top: 1, 2.5, 5, 7.5 and 10 sccm. Literature binding energies for B-W [33] and O-W [33] are indicated by the vertical dotted lines in a) and b), respectively.

The C *1s* peaks in Fig. 2b are located at 283.4 eV, which is close to the reported binding energy between 283.4 and 283.6 eV reported for sputtered $W_2C$ films with a hexagonal crystal structure by Greczynski *et al.* [39] as well as Liu *et al.* for W-B-C films [25]. The average of these measurements is indicated by a vertical dotted line at 283.5 eV in Fig. 2b. and potential C-W* bonding [40]. In addition, the C *1s* spectra show no evidence of C-O bonding in the deposited films as no peak is visible at ~292 eV [34].

The B *1s* peaks in Fig. 3a, centered at 187.9 eV, are of low intensity for films deposited with 1 and 2.5 sccm TMB. This binding energy is identical to the measured B-W binding energy for $W_2B_5$ [34] and W-B-C films deposited by Alishahi *et al.* [26], and close to the 188.1 eV reported by Liu *et al.* for W-B-C films [25]. There is no indication of B-B or B-O bonding in the films as no B *1s* peaks are visible at 189.4 eV [34] or ~192-193 [34], respectively.

The O *1s* peaks in Fig. 3b display low intensities and asymmetric tails towards higher binding energy. The peak position located at ~530.6 eV is identical to that determined for $WO_3$ and that is indicated by a vertical dotted line in the spectra [34]. No corresponding peaks from oxygen bonding are observed in the W *4f*, C *1s*, and B *1s* spectra.

The binding energies for the W *4f$_{7/2}$*, W *4f$_{5/2}$*, C *1s*, B *1s*, and O *1s* peaks remain constant for films deposited at RT, 100, 200, 300, 400, and 600 °C and using 10 sccm TMB in the plasma compared to the films deposited at 500 °C with 10 sccm TMB. For the film deposited at RT conditions this is a support of the strong driving force for W to form carbides and borides. When increasing the temperature to 700 °C, Si is detected by XPS.

## 3.2. Phase distribution in the deposited films

Fig. 4a shows θ/2θ diffractograms from W-B-C films deposited with TMB flows of, from bottom to the top: 1, 2.5, 5, 7.5, and 10 sccm at 500 °C. The peaks in the diffractogram originating from the W-B-C films are broad and generally of low intensities, indicating a nanocrystalline microstructure. In addition, there are peaks from the Si(100)substrate dominated by the 400 peak at 2θ ≈ 69° and the much weaker 600 peak at 2θ ≈ 117°. For some of the films the "forbidden" 200 peak at 2θ ≈ 33° [41] is visible as well as an artefact from the applied diffractometer at 2θ ≈ 17°. The film deposited with 1 sccm TMB in the plasma show peaks at the 2θ angles ~40°, ~58°, ~87°, and ~101°, corresponding to the 110, 200, 220, and 310 peaks from W [42] and with the 2θ angles for W determined in literature indicated by dotted vertical lines in Fig. 4a. There is a peak at 2θ ≈ 43° from the cubic $WC_{1-x}$ phase, which is the 200 peak as calculated from the *a*-axis lattice parameter 4.2355 Å [43] using Cu-$K_\alpha$ radiation instead of Co-$K_\alpha$ as in ref. [43]. The position of the 200 peak is indicated as a dotted line in the diffractogram. The presence of $WC_{1-x}$ in the film shows a strong driving force towards carbide formation at low flows of TMB and is supported by the XPS analysis in section 3.1. Increasing the flow of TMB to 2.5 sccm broadens and reduces the number of peaks in the diffraction pattern, while the peaks from the W phase are retained. Increasing the TMB flow further to 5 sccm results in two visible peaks originating from $WC_{1-x}$ corresponding to the 200 and 400 peaks at 2θ angles of ~42° and ~95°, respectively. A strong peak at a 2θ ~40° is reported in the publications by Alishahi *et al.* [26] and Debnárová *et al.* [27] for more boron and carbon rich W-B-C films and with a second peak at 2θ ~70° [26] that is not visible for our deposited films. From the diffractogram in Fig. 4a there is no evidence of crystalline δ-WC that should exhibit strong peaks at 2θ ≈ 36° 10$\bar{1}$0 and 2θ ≈ 48° 10$\bar{1}$1 [44], or crystalline tungsten borides such as $WB_2$ with $AlB_2$ type-structure (Strukturbericht designation C32) with strong peaks at 2θ ≈ 29° 0001 and 2θ ≈ 46° 10$\bar{1}$1 [45], or hexagonal $W_2B_5$ with strong peaks at 2θ ≈ 25° 0004 and 2θ ≈ 35 10$\bar{1}$0 and 10$\bar{1}$1 [46]. For tetragonal WB there is an overlap between two strong peaks, 105 and 112, with the $WC_{1-x}$ 200 peak, but with no matching peaks expected at 2θ ≈ 94°, and where a strong 103 peak at 2θ ≈ 33° is absent [47]. To conclude, the θ/2θ diffraction shows no evidence for either crystalline δ-WC or tungsten borides, and where the $WC_{1-x}$ phase is the preferred crystalline phase in agreement with what has previously been reported for sputtered WC films [11] [14] [17] [18] [21] [23]. The relatively strong intensities of the 200 and 400 peaks in the diffractogram suggests that the $WC_{1-x}$ film has corresponding preferred crystallographic orientation *i.e.*, 100-oriented.

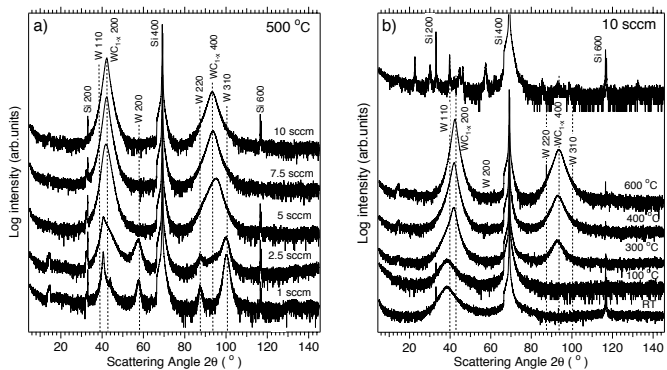

**Figure 4:** (a) X-ray θ/2θ diffractograms recorded from W-B-C films deposited at 500 °C and with TMB flows from bottom to the top: 1, 2.5, 5, 7.5 and 10 sccm. (b) X-ray θ/2θ diffractograms recorded from W-B-C films deposited from bottom to the top at: RT, 100, 300, 400, 600, and 700 °C with 10 sccm TMB in the plasma.

To investigate the film orientation with respect to the substrate, we applied pole figure measurements, see Fig. 5. As observed, the monitored {200} pole displays a point of high intensity centered in the middle of the figure where ψ ≈ 0°, i.e. in the growth direction. The investigated {220} and {222} pole figures have high intensity rings located at ψ ≈ 45° and ψ ≈ 54.7°, respectively, which is expected for a 100-oriented fiber-textured film [48].

Additional pole figure measurements at the expected positions of {200}, {211}, {220} and, {310} in a carbon supersaturated bcc-structured α-W phase (not shown), as observed by *e.g.* Pauleau for W-C coatings containing less than 25 at.% C refs. [49][51], was conducted to eliminate the possibility of such a phase being present in our films.

At TMB flows of 7.5 and 10 sccm the 200 and 400 peaks increase in intensities and there are no indications of crystalline borides in the films even though the films contain 6-7 at.% B. This suggests that W has a stronger driving force to form carbides compared to borides in reactive sputtering with TMB, but where a solid solution of B in the $WC_{1-x}$ phase cannot be excluded. This is supported by XPS in Fig. 3a that shows W-B bonding.

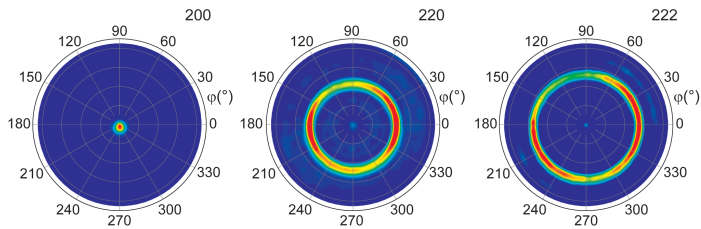

**Figure 5:** Pole figure measurements of the 200, 220, and 222 reflections for a W-B-C film deposited at 500 °C and with a 10 sccm TMB flow showing a 100-oriented fiber texture.

Fig. 4b shows θ/2θ diffractograms from films deposited at deposited at from bottom to the top: RT, 100, 300, 400, 600, and 700 °C and with 10 sccm TMB in the plasma. Already for growth at RT conditions there is a broad peak centered at 2θ≈38° beside the peaks previously described from the Si(100) substrate in Fig. 4a. The peak is centered at a lower diffraction angle compared to the literature values for the W 110 peak at 2θ ≈ 40.26° [42] and the $WC_{1-x}$ 200 peak at 2θ ≈ 42.66°, i.e. at the right-hand side of the peak and with the literature values determined for W and $WC_{1-x}$ indicated as dotted line in the diffractogram. The peak position is close to orthorhombic [51] and hexagonal [52] $W_2C$ at with peaks 2θ ≈ 38°, but where the absence of other peaks makes phase identification uncertain. In addition, we note the possibility of a supersaturated solid solution of C in α-W (bcc) [21][49][50], and/or an amorphous phase content.

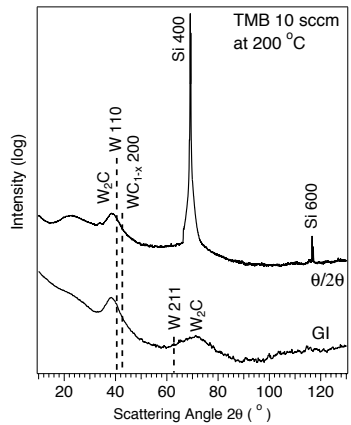

**Figure 6:** θ/2θ and grazing incidence (GI) (θ=2°) XRD measurements of a W-B-C film deposited at 200 °C and with a 10 sccm TMB flow. The θ/2θ diffractogram is vertically shifted for clarity.

The peak remains centered at 2θ ≈ 38° when the substrate temperature is increased to 100 °C and 200 °C (not shown). For additional investigation of the phase distribution in these films, we applied diffraction in GI geometry.

Figure 6 shows GI-XRD measurement of a W-B-C film deposited at 200 °C and with a 10 sccm TMB flow, revealing two broad structures centered around 2θ ≈ 38° and 2θ ≈ 72°. These angles are close to expected positions for high intensity peaks of both orthorhombic [51] and hexagonal [52] $W_2C$. The corresponding θ/2θ diffractogram in Fig. 6 provides no help for determining the $W_2C$ polytype as the substrate 400 diffraction peak overlap with the peaks at 2θ ≈ 72°. For the broad peak at 2θ ≈ 38°, both geometries suggest a phase mixture with $W_2C$, possible supersaturated solid solution of C in α-W (bcc) or $WC_{1-x}$ contributions for films deposited at 100 °C (not shown) and 200 °C with 10 sccm TMB in the plasma.

At a deposition temperature of 300 °C the peak at 2θ ≈ 38° in Fig. 4b shifts to a higher diffraction angle of ~42°, and a second peak is found at ~92°. These are the 200 and 400 peaks from $WC_{1-x}$ and indicate an oriented growth to the Si(100) substrate already at 300 °C. Increasing the growth temperature to 400, 500 (in Fig. 4a) and 600 °C results in increasing intensities for the $WC_{1-x}$ 200 and 400 peaks. However, at 700 °C the diffraction pattern completely changes seen from the number of peaks and their positions. This is attributed to a reaction with the substrate, where Si is diffusing into the film and crystalline tetragonal $WSi_2$ [53] is formed. The $WSi_2$

peaks are seen from increasing 2θ: 002, 101, 110, 103, 112, 200, and 202/114 at 2θ ≈ 63°, and where the intensity distribution for the peaks indicate a random orientation for the film. Silicide formation during sputtering of W on Si(100) substrates has previously been reported following annealing at 1073 K (800 °C) [54]. The diffraction peaks from $WSi_2$ gain in intensities at 800 °C, but reduces in intensities at 900 °C to become extinct, *i.e.* an X-ray amorphous film was formed (not shown). For this film, XPS measurements show a surface composition with a Si content of ~96 at.%.

*3.3. Morphology of the films*

Figure 7 shows SEM images obtained from films deposited at TMB flows of 1, 2.5, 5, and 10 sccm at 500°C in (a) and films deposited at 200, 400, 600, and 900 °C and with 10 sccm TMB in the plasma in (b). The films deposited with different TMB content in the plasma exhibit thicknesses in the range ~800 to ~1000 nm corresponding to a deposition rate of 160 to 200 nm/min and with higher thicknesses for the films deposited with 5 and 10 sccm TMB in the plasma. The microstructure changes from broken columns, as in the film deposited with 1 sccm TMB in the plasma to fine-grained in the film grown with 10 sccm. A fine-grained microstructure for $WC_{1-x}$ films finds support from literature, where $WC_{1-x}$ films are often described as being nanocrystalline/ fine-grained [14] [23]. In addition, Mitterer *et al.* showed that alloying TiC with B favor a fine-grained microstructure [55]. The surface structure of the films is generally smooth, which is supported from ocular inspection of films that are the silver-white with a metallic luster. Furthermore, the $WC_{1-x}$ films deposited at 400 and 600 °C in Fig. 7 also exhibit a fine-grained microstructure c.f. Fig. 4b. Decreasing the deposition temperature to

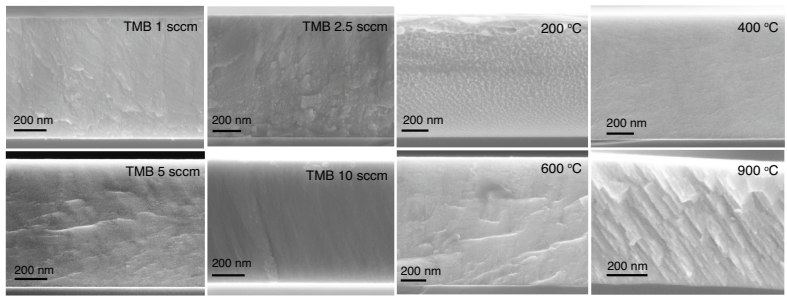

**Figure 7:** (a) SEM images obtained from W-B-C films deposited at 500°C and with TMB flows of 1, 2.5, 5, and 10 sccm. (b) SEM images obtained from W-B-C films deposited at 200, 400, 600, and 900 °C with 10 sccm TMB in the plasma.

200 °C results in a more glass-like microstructure possibly associated with the growth of $W_2C$, see Fig. 4b and Fig. 7, but where the film maintain its silver-white color and metallic luster. Increasing the deposition temperature to 900 °C results in a fine-grained microstructure and with an increased variation of the thickness in the film as well as increased surface roughness. This is accompanied by a change to a greyish colored surface. Our XRD analysis in section 3.2 explained this being the result of a reaction between the film and the Si(100) substrate. Such reaction between the film and the substrate is likely to compromise the properties of the films such as the H and $E_r$.

*3.4 Mechanical properties*

The H and $E_r$ of the films were assessed by nanoindentation with films deposited at TMB flows of 1, 2.5, 5, 7.5, and 10 sccm at 500 °C in Fig. 8a and films deposited at 200, 300, 400, 500, 600, 700, 800, and 900 °C and with 10 sccm TMB in the plasma in Fig. 8b. The measured H and $E_r$ values show no clear trends for the films deposited at different TMB flows in the plasma. In contrast, both the H and the $E_r$ decrease markedly by a factor of three to four for the films deposited at 700, 800, 900 °C that is accompanied by an increased spread between the individual measurements during nanoindentation. This is a consequence of the reaction between the Si(100) substrate and the film, resulting in formation of $WSi_2$ at 700 and 800 °C or a Si rich surface at 900 °C as determined from our XRD and XPS measurements and where the reaction also yields an increased surface roughness seen from the loss of the silver-white color and metallic luster. The H varies in the range 23 to 31 GPa for films deposited with different TMB flows and at temperatures ≲ 600 °C; a closer inspection shows the lowest H values for the films deposited at temperatures of 200, 300 and 400 °C, with H = 23, 25, and 24 GPa, respectively.

Our H values are comparable or even slightly higher than the reported HV of δ-WC with 22 GPa for a 0001-oriented single-crystal [2] as well as to those reported for more carbon and boron rich W-B-C films by Alishahi *et al.* [26] and Debnárová *et al.* [27] with values in the range ~24 to ~29 GPa and ~23 GPa, respectively. In contrast, our H values are lower than the ~37 to ~47 GPa reported by Su *et al.* [24] for δ-WC films alloyed with ~4 to ~7 at. N as well as the ~45 GPa determined by Liu *et al.* for W-B-C films reactively sputtered from a $WB_2$ target [25]. The H values measured by Su *et al.* support the superior mechanical properties of δ-WC films compared to $WC_{1-x}$ films in particularly when δ-WC is alloyed with N. For the W-B-C films reported by Liu *et al.* it is important to firstly note the higher boron content in their films with ~50 at% compared to our films with 6-8 at.% B. Secondly, the phase distribution in their films is different with peak(s) from crystalline $WB_2$ with C32 crystal structure [25], while diffractograms recorded from our films show crystalline $WC_{1-x}$ with a possible solid solution of B. This suggests less favorable hardness in $WC_{1-x}$ films compared to $WB_2$ films as H values of ~40 GPa have been measured for monolithic $WB_2$ films [45], but where bulk $W_2B_5$ is reported to exhibit a HV value of 26.1 GPa [2], *i.e.* comparable to our films.

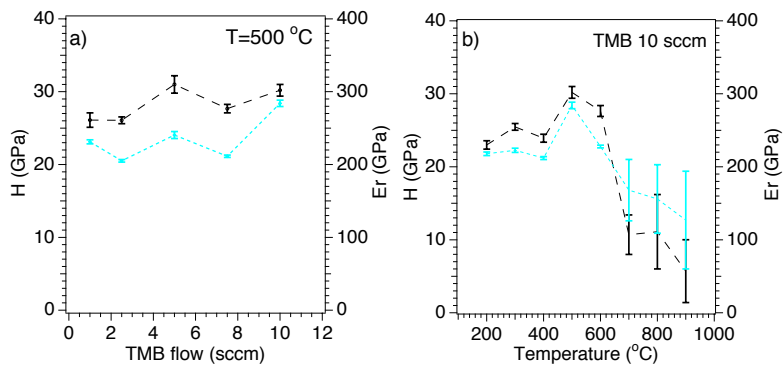

**Figure 8:** Mechanical properties, H as darker dashed lines and $E_r$ lighter dotted lines, for W-B-C films deposited at 500°C and with TMB flows of 1, 2.5, 5, 7.5, and 10 sccm in (a) and for films deposited at 200, 300, 400, 500, 600, 800, and 900 °C and with 10 sccm TMB in the plasma in (b). The lines are guides for the eye.

Furthermore, the measured H values are higher than the ~17 GPa obtained by Palmquist *et al.* for polycrystalline $W_2C$ and $WC_{1-x}$ films with ~22 at.% C [14]. From this study we note the higher H value of 34.5 GPa measured for an epitaxial W film with 7 at.% C, suggesting that the H depends on film orientation. Indentation, including HV measurements, of more carbon rich films show scattered values of ~17 GPa [23] and 3500 to 4500 $HV_{0.025}$ [17] corresponding to ~32 to 41 GPa using the correction factors described in the Experimental details section, but where the authors acknowledge that the measured H values are likely affected by the low thickness of the investigated films [17]. For sputtered WC/a-C:H films, Drábik *et al.* presents H values in the region ~13 to 15 GPa [20], showing that growth of nanocomposites results in reduced hardness in the W-C-H system. Thus, the measurements show that our reactively sputtered W-rich films exhibit comparable or even superior H values to bulk δ-WC as well as to more carbon and/or boron rich films.

The $E_r$ measured in our films follow the same trends previously described for hardness, see Fig. 8a and 8b, with the highest value of ~ 280 GPa for the film deposited with 10 sccm TMB at 500 °C and a lowest value of ~130 GPa for the film deposited at 900 °C. Similar as for the hardness, our $E_r$ ~ 280 GPa value is similar to the $E_r$ ~ 300 GPa reported by Debnárová *et al*. [27] but lower than the ~ 450 GPa reported in the studies by Alishahi *et al*. [26]. Furthermore, for $W_2C$ and $WC_{1-x}$ films containing 35 at.% C Palmquist *et al*. [14] measured an E-moduli of ~ 450 GPa, i.e. similar to that of Alishahi *et al*. We attribute the lower $E_r$ values of our films compared to Alishahi *et al.* and Palmquist *et al* to the lower carbon and boron contents in our films with ~ 18 at.% and ~ 6 at.%, respectively.

## 4. Conclusions

Reactive sputtering of W in Kr/TMB plasmas results in growth of W-rich 100-oriented $WC_{1-x}$ with a potential boron solid solution. The applied TMB flow with ~93 at.% W at 1 sccm and ~72 at.% W at 10 sccm determines the metal content for films deposited ≤ 600 °C. The C and B contents in such films reach a maximum of ~18 at.% and ~7 at.% respectively and where both elements are chemically bonded to W. Growth conditions with TMB flows ≥ 5 sccm and temperatures ≤ 600 °C result in deposition of crystalline and 100-oriented $WC_{1-x}$ with a possible solid solution of B whereas lower TMB flows give W films and temperatures below 300 °C yield $W_2C$ films. For films deposited at different TMB flows and temperatures ≤ 600 °C, nanoindentation shows hardness values in the range ~23 to ~31 GPa and reduced elastic moduli between ~220 and 280 GPa. Temperatures above 600 °C leads to the formation of $WSi_2$ as introduced by a reaction between the film and the Si(100) substrate. This reaction changes the color of the films from silver-white with metallic luster to greyish and with increased surface roughness as well as reduces the H and $E_r$ of the films by a factor of three to four.

**Acknowledgements**
The research leading to these results has received funding from the Swedish Government Strategic Research Area in Materials Science on Advanced Functional Materials at Linköping University (Faculty Grant SFO-Mat-LiU No. 2009-00971). MM acknowledges financial support from the Swedish Energy Research (no. 43606-1), VINNOVA ( 2018-04410, 2018-04417), the Swedish Foundation for Strategic Research (SSF) (no. RMA11-0029) through the synergy grant FUNCASE and the Carl Tryggers Foundation (CTS16:303, CTS14:310). GG thanks the Knut and Alice Wallenberg Foundation Scholar Grant KAW2016.0358, the VINN Excellence Center Functional Nanoscale Materials (FunMat-2) Grant 2016-05156, and the Åforsk Foundation Grant 16-359. Harri Savimäki is acknowledged for construction of the gas handling system.

## References
[1] T. Massalski, J. Murray, L. Bennett and H. Baker, Binary alloy phase diagrams, Metals Park, OH: American Society for Metals, 1986.
[2] H. O. Pierson, Handbook of refractory metal carbides and nitrides: properties, characteristics, processing and applications, NJ: Noyes publications, 1996.
[3] N. Lundberg, M. Östling, C.-M. Zetterling, P. Tägtström and U. Jansson, CVD-Based Tungsten Carbide Schottky Contacts to 6H-SiC for Very High-Temperature Operation, J. Elec. Mat., 29 (2000) 372-375.
[4] T. Takahashi and H. Itoh, Chemical vapor deposition of tungsten carbide dendrites, J. Cryst. Growth, 12 (1972) 265-271.

[5] M. Katoh and H. Kawarada, Heteroepitaxial growth of tungsten carbide films on W(110) by plasma-enhanced chemical vapor deposition, Jpn. J. Appl. Phys. 34 (1995) 3628-3630.

[6] M. Fitzsimmons and V. K. Sarin, Comparion of WCl6-CH4-H2 and WF6-CH4-H2 systems for growth of WC coatings, Surf. Coat. Technol., 76-77 (1995) 250-255.

[7] P. Tägtström, H. Högberg, U. Jansson and J.-O. Carlsson, Low pressure CVD of tungsten carbides, Journal de Physique IV, Colloque C5, supplement au Journal de Physique II, C5 (1995) 967-974.

[8] H. Högberg, P. Tägtström, J. Lu and U. Jansson, Chemical vapour deposition of tungsten carbides on tatalum and nickel substrates, Thin Solid Films, 272 (1996) 116-123.

[9] W. S. Williams, Physics of transition metal carbides, Materials Science and Engineering A, 105-106 (1988) 1-10.

[10] A. K. Taylor, High speed friction and wear characteristics of vapor deposited coatings of titanium carbide and tungsten carbide plus cobalt in a vacuum environment, Thin Solid Films 40 (1977) 189-200.

[11] K. Fuchs, P. Rödhammer, E. Bertel, F. P. Netzer and E. Gornik, Reactive and non-reactive high rate sputter deposition of tungsten carbides, Thin Solid Films, 151 (1987) 383-395.

[12] E. C. Weigert, M. P. Humbert, Z. J. Mellinger, Q. Ren, T. P. Beebe Jr., L. Bao and J. G. Chen, Physical vapor deposition synthesis of tungsten carbide monocarbide (WC) thin films on different carbon substrates, J. Vac. Sci. Technol. A, 26, (2008) 23-28.

[13] K. A. Taylor, Ancillary properties of vapor-deposited carbide coatings, thin solid films, 40 (1977) 201-209.

[14] J.-P. Palmquist, Z. Czigany, M. Odén, J. Neidhardt, L. Hultman and U. Jansson, Magnetron sputtered W-C films with C60 as carbon source, Thin Solid Films 444 (2003) 29-37.

[15] J.-P. Palmquist, Z. Czigány, L. Hultman and U. Jansson, Epitaxial growth o tungsten carbide films using C60 as carbon precursor, J. Cryst. Growth, 259 (2003) 12-17.

[16] P. K. Srivastava, V. D. Vankar and K. L. Chopra, High rate reactive magnetron sputtered tungsten carbide films, J. Vac. Sci. Technol. A, 3 (1985) 2129-2134.

[17] G. Keller, R. Erz, I. Barzen, M. Weiler, K. Jung and H. Ehrhardt, Mechanical properties, structure and composition of ion-plated tungsten carbide films, Vacuum, 41 (1990) 1294-1296.

[18] G. Keller, I. Barzen, R. Erz, W. Dötter, S. Ulrich, K. Jung and H. Ehrhardt, Crystal structure, morphology and composition of magnetron sputtered tungsten carbide films, Fresenius J. Anal. Chem. 341 (1991) 349-351.

[19] C. Rebholz, J. M. Schneider, A. Leyland and A. Matthews, Wear behavior of carbon-containing tungsten coatings prepared by reactive magnetron sputtering, Surf. Coat. Technol., 112 (1999) 85-90.

[20] M. Drábik, M. Truchlý, V. Ballo, T. Roch, L. Kvetková and P. Kúš, Influence of substrate material and its plasma pretreatment on adhesion and properties of WC/a-C:H nanocomposite coatings deposited at low temperatures, Surf. Coat. Technol. 333 (2018) 138-147.

[21] P. Gouy-Pallier and Y. Pauleau, Tungsten and tungsten-carbon thin films deposited by magnetron sputtering, J. Vac. Sci. Technol. A, 11 (1993) 96-102.

[22] P. D. Rack, J. J. Peterson, J. Li, A. C. Geiculescu and H. J. Rack, Thin film growth of reactive sputter deposited tungsten-carbon thin films, J. Vac. Sci. Technol. A 19 (2001) 62-65.

[23] N. Radić, B. Gržeta, O. Milat, J. Ivkov and M. Stubičar, Tungsten-carbon films prepared by reactive sputtering from argon-benzene discharges, Thin Solid Films 320 (1998) 192-197.

[24] Y. D. Su, C. Q. Hu, C. Wang, M. Wen and W. T. Zheng, Realatively low temperature synthesis of hexagonal tungsten carbide films by N doping and its effect on the preferred orientation, phase transition and mechanical properties, J. Vac. Sci. Technol. A, 27 (2009) 167-173.

[25] Y. M. Liu, Z. L. Pei, J. Gong and C. Sun, Effect of carbon content on microstructures, mechanical properties and tribological properties and thermal stability in WBC films, Surf. Coat. Technol., 291 (2016) 276-285.

[26] M. Alishahia, S. Mirzaei, P. Souček, L. Zábranský, V. Buršíková, M. Stupavská, V. Peřina, K. Balázsi, Z. Czigány and P. Vašina, Evolution of structure and mechanical properties of hard yet fracture resistant W-B-C coatings with varying C/W ratio, Surf. Coat. Technol. 340 (2018) 103–111.

[27] S. Debnárová, P. Souček, P. Vašina, L. Zábranský, V. Buršíková, S. Mirzaei and Y. T. Pei, The tribological properties of short range ordered W-B-C protective coatings prepared by pulsed magnetron sputtering, Surf. Coat. Technol. 357 (2019) 364–371.

[28] J. S. Lewis, S. Vaidyaraman, W. J. Lackey, P. K. Agrawal, G. B. Freeman and E. K. Barefield, Chemical vapor deposition of boron-carbon films using organometallic reagents, Mat. Lett. 27 (1996) 327-332.

[29] M. Chubarov, H. Pedersen, H. Högberg, V. Darakchieva, J. Jensen, P. O. Å. Persson and A. Henry, Epitaxial CVD growth of sp2-hybridized boron nitride using aluminium nitride as buffer layer, Phys. Stat. Solidi RRL 5 (2011) 397-399.

[30] M. Chubarov, H. Pedersen, H. Högberg, J. Jensen and A. Henry, Growth of high quality epitaxial rhombohedral boron nitride, Cryst. Growth Des. 12 (2012) 3215-3220.

[31] M. Chubarov, H. Pedersen, H. Högberg, Z. Czigany and A. Henry, Chemical vapour deposition of epitaxial rhombohedral BN thin films on SiC substrates, Cryst. Eng. Comm. 16 (2014) 5430-5436.

[32] H. Högberg, J. Birch, M. P. Johansson and L. Hultman, Deposition of epitaxial transition metal carbide films and superlattices by simultaneous direct current metal magnetron sputtering and C60 evaporation; J. Mater Res. **16**, (2001) 633-643.

[33] S. Hüfner, Photoelectron spectroscopy: Principles and applications, 3rd edn., Berlin, Germany: Springer-Verlag, (2010).

[34] J. F. Moulder, W. F. Stickle, P. E. Sobol and K. D. Bomben, Handbook of X-ray photoelectron spectroscopy. Perkin-Elmer Corporation, Eden Prairie, USA, (1992).

[35] W. C. Oliver and G. M. Pharr, W. Oliver and G. Pharr, Measurement of hardness and elastic modulus by instrumented indentation: Advances in understanding and refinnements to methodology, J. Mater. Res. 19 (2004) 3-20.

[36] E. Broitman, Indentation hardness measurements at macro-, micro-, and nanoscale: a critical overview, Trib. Lett. 65, (2017) 23.

[37] ISO 14577-4:2015, Metallic materials - Instrumented indentation test for hardness and materials parameters, International Organization for Standardization, Geneva, Switzerland (2015).

[38] Thompson, A. C.; Vaughan, D., X-ray Data Booklet, Lawrence Berkeley National Laboratory University of California Berkeley, CA 94720, (2001). http://xdb.lbl.gov/.

[39] G. Greczynski, D. Primetzhofer and L. Hultman, Reference binding energies of transition metal carbides by core-level x-ray photoelectron spectroscopy free from Ar+ etching artefacts, Appl. Surf. Sci. 436 (2018) 102–110.

[40] M. Magnuson, E. Lewin, L. Hultman and U. Jansson, Electronic structure and chemical bonding of nanocrystalline-TiC/amorphous-C nanocomposites, Phys. Rev. B 80 (2009) 235108.

[41] P. Zaumseil, High-resolution characterization of the forbidden Si 200 and Si 222 reflections, J. Appl. Cryst. 48 (2015) 528-532.

Thin Solid Films **688**, 137384 (2019). DOI: 10.1016/j.tsf.2019.06.034

[42] JCPDS - International Centre for Diffraction Data of W, Ref ID: 00-04-806.

[43] JCPDS - International Centre for Diffraction Data WC1-x, Ref ID: 00-20-1316.

[44] JCPDS - International Centre for Diffraction Data of delta-WC, Ref ID: 00-20-1047.

[45] Y. M. Liu, R. Q. Han, F. Liu, Z. L. Pei and C. Sun, Sputtering gas pressure and target power dependence on the microstructure and properties of DC-magnetron sputtered AlB2-type WB2 films, J. Alloy Comp. 703 (2017) 188-197.

[46] JCPDS - International Centre for Diffraction Data of hexagonal W2B5, Ref ID: 38-1365.

[47] JCPDS - International Centre for Diffraction Data of tetragonal delta-WB, Ref ID: 35-0738.

[48] M. Birkholz, Thin Film Analysis by X-Ray Scattering, Weinheim: Wiley-VCH Verlag GmbH & Co., Germany, 2006.

[49] Y. Pauleau and Ph. Gouy-Pailler, Very hard solid-solution-type tungsten-carbon coatings deposited by reactive magnetron sputtering, Mater. Lett. 13 (1992) 157-160.

[50] Y. Pauleau and Ph. Gouy-Pailler, Characterization of tungsten-carbon layers deposited on stainless steel by reactive magnetron sputtering, J. Mater. Res. 7 (1992) 2070-2079.

[51] JCPDS - International Centre for Diffraction Data orthorombic W2C, Ref ID: 20-1315.

[52] JCPDS - International Centre for Diffraction Data, hexagonal W2C, Ref ID: 35-0776.

[53] JCPDS - International Centre for Diffraction Data of WSi2, Ref ID: 11-0195.

[54] M. O. Aboelfotoh, Electrical characteristics of W-Si(100) Schottky barrier junction, J. Appl. Phys. 66 (1989) 262-272.

[55] C. Mitterer, P. H. Mayrhofer, M. Beschlisser, P. Losbichler, P. Warbichler, F. Hofer, N. P. Gibson, W. Gissler, H. Hruby, J. Musil and J. Vlček, Microstructure and properties of nanocomposite Ti-B-N and Ti-B-C coatings, Surf. Coat. Technol. 120-121 (1999) 405-411.